\documentclass[aps,prl,twocolumn,superscriptaddress,floatfix]{revtex4}

\usepackage{graphics}
\usepackage{epsfig}
\newcommand{\abu}
{\affiliation{Department of Physics, Boston University, Boston, MA
02215, USA}}
\newcommand{\aucb}
{\affiliation{Department of Physics, University of California,
Berkeley, CA 94720, USA}}
\newcommand{\auiuc}
{\affiliation{Department of Physics, University of Illinois at
Urbana-Champaign, Urbana, IL 61801, USA}}
\newcommand{\ajmu}
{\affiliation{Department of Physics, James Madison University,
Harrisonburg, VA 22807, USA}}
\newcommand{\aky}
{\affiliation{Department of Physics and Astronomy, University of
Kentucky, Lexington, KY 40506, USA}}
\newcommand{\akvi}
{\affiliation{Kernfysisch Versneller Instituut, Groningen
University, NL 9747 AA Groningen, The Netherlands}}
\newcommand{\atriumf}
{\affiliation{TRIUMF, Vancouver, BC, V6T 2A3, Canada}}
\begin{document}

\title{Improved Measurement of the Positive Muon Lifetime and \\ Determination of the Fermi Constant}

\author{D.B.\,Chitwood}
\auiuc
\author{T.I.\,Banks}
\aucb
\author{M.J.\,Barnes}
\atriumf
\author{S.\,Battu}
\aky
\author{R.M.~Carey}
\abu
\author{S.\,Cheekatmalla}
\aky
\author{S.M.\,Clayton}
\auiuc
\author{J.\,Crnkovic}
\auiuc
\author{K.M.\,Crowe}
\aucb
\author{P.T.\,Debevec}
\auiuc
\author{S.\,Dhamija}
\aky
\author{W.\,Earle}
\abu
\author{A.\,Gafarov}
\abu
\author{K.\,Giovanetti}
\ajmu
\author{T.P.\,Gorringe}
\aky
\author{F.E.\,Gray}
\auiuc \aucb
\author{M.\,Hance}
\abu
\author{D.W.\,Hertzog}
\auiuc
\author{M.F.\,Hare}
\abu
\author{P.\,Kammel}
\auiuc
\author{B.\,Kiburg}
\auiuc
\author{J.\,Kunkle}
\auiuc
\author{B.\,Lauss}
\aucb
\author{I.\,Logashenko}
\abu
\author{K.R.\,Lynch}
\abu
\author{R.\,McNabb}
\auiuc
\author{J.P.\,Miller}
\abu
\author{F.\,Mulhauser}
\auiuc
\author{C.J.G.~Onderwater}
\auiuc \akvi
\author{C.S.~\"{O}zben}
\auiuc
\author{Q.~Peng}
\abu
\author{C.C.~Polly}
\auiuc
\author{S.\,Rath}
\aky
\author{B.L.\,Roberts}
\abu
\author{V.\,Tishchenko}
\aky
\author{G.D.\,Wait}
\atriumf
\author{J.\,Wasserman}
\abu
\author{D.M.\,Webber}
\auiuc
\author{P.\,Winter}
\auiuc
\author{P.A.\,\.{Z}o{\l}nierczuk}\aky

\collaboration{MuLan Collaboration}

\begin{abstract}
The mean life of the positive muon has been measured to a precision
of 11~ppm using a low-energy, pulsed muon beam stopped in a
ferromagnetic target, which was surrounded by a scintillator
detector array. The result, $\tau_{\mu} = 2.197\,013(24)~\mu$s, is
in excellent agreement with the previous world average. The new
world average $\tau_{\mu} = 2.197\,019(21)~\mu$s determines the
Fermi constant $G_F = 1.166\,371(6) \times 10^{-5}$~GeV$^{-2}$
(5~ppm). Additionally, the precision measurement of the positive
muon lifetime is needed to determine the nucleon pseudoscalar
coupling ${\textsl g}^{}_P$.

\end{abstract}

\maketitle

The predictive power of the standard model relies on precision
measurements of its input parameters. Impressive examples include
the fine-structure constant~\cite{gerry}, the $Z$~mass~\cite{ewwg},
and the Fermi constant~\cite{vanRitbergen:All}, having relative
precisions of 0.0007, 23, and 9~ppm, respectively.

The Fermi constant $G_F$ is related~\cite{vanRitbergen:All} to the
electroweak gauge coupling $g$ by
\begin{equation}
\frac{G_F}{\sqrt{2}} = \frac{g^2}{8 M_W^2}\left( 1 + \Delta r
\right), \label{eq:Gftog}
\end{equation}
where $\Delta r$ represents the weak-boson-mediated tree-level and
radiative corrections, which have been computed to second
order~\cite{Awramik:TwoLoopMuonLifetime}.  Comparison of the Fermi
constant extracted from various measurements stringently tests the
universality of the weak interaction's
strength~\cite{fermiconstants}.

The most precise determination of $G_F$ is based on the mean life of
the positive muon,  $\tau_{\mu}$.  It has long been known that in
the Fermi theory, the QED radiative corrections are finite to first
order in $G_F$ and to all orders in the electromagnetic coupling
constant, $\alpha$~\cite{BermanSirlin}. This provides a
framework~\cite{vanRitbergen:All} for extracting $G_F$ from
$\tau_{\mu}$,
\begin{equation}
\frac{1}{\tau_\mu}= \frac{G_F^2 m_\mu^5}{192 \pi^3} \left( 1+\Delta
q \right), \label{eq:muondecay}
\end{equation}
where $\Delta q$ is the sum of phase space and both QED and hadronic
radiative corrections, which have been known in lowest-order since
the 1950s~\cite{oldQED}.  Relation~\ref{eq:muondecay} does not
depend on the specifics of the underlying electroweak model.

Until recently, the uncertainty on extracting $G_F$ from $\tau_\mu$
was limited by the uncertainty in higher-order QED corrections,
rather than by measurement. In 1999, van Ritbergen and Stuart's
calculation of the second-order QED
corrections~\cite{vanRitbergen:All} reduced the relative theoretical
uncertainty in the determination of $G_F$ to less than 0.3~ppm. The
dominant uncertainty is currently from $\tau_\mu$, which motivates
this work. While the final goal of our experiment is a 1~ppm
uncertainty on $\tau_{\mu}$, we report here a result having a
precision of 11~ppm---2.5 times better than any previous
measurement~\cite{lifetimes}---based on data obtained in 2004, the
commissioning run period.

\begin{figure}
\begin{center}
\includegraphics[width=\columnwidth]{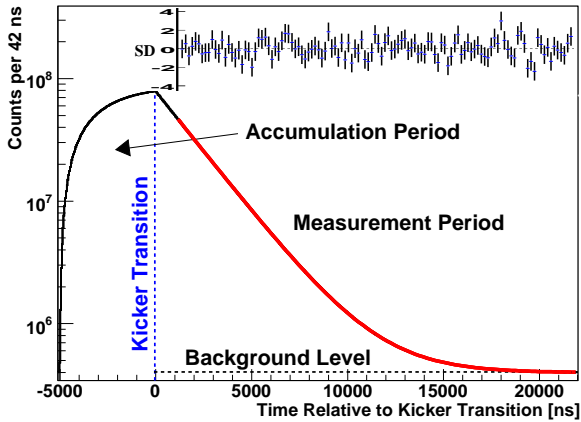}
\end{center}
\caption{Data from a subset of the MuLan detectors, illustrating the
accumulation and measurement periods and the background mainly
caused by incomplete extinction. The fit region used for the whole
data set is indicated as a thick red line.  The residuals divided by
their uncertainty (i.e, in standard deviations) are shown in the
upper inset panel.\label{fig:cycleplot}}
\end{figure}

The experimental design is conceptually simple. Longitudinally
polarized muons from the $\pi$E3 beamline at the Paul Scherrer
Institute are stopped in a thin ferromagnetic target. A
fast-switching kicker imposes a cycle on the continuous beam,
consisting of a $5~\mu$s ``beam-on'' period of stopped-muon
accumulation, $T_A$, followed by a $22~\mu$s ``beam-off''
measurement period, $T_M$.  The muon decay positrons are detected in
a scintillator array which surrounds the target. A decay time
spectrum for a subset of the detectors is shown in
Fig.~\ref{fig:cycleplot}. The background level is indicative of the
``extinction'' of the beam during $T_M$, caused by the kicker.

The beamline is tuned to transport 28.8~MeV/$c$  muons from pions
that decay at rest near the surface of the production target. Two
opposing $60^{\circ}$ bends raise the beam from ground level to the
experimental area, where it is directed parallel to the optic axis
through an $\vec{E}\times\vec{B}$ velocity separator that removes
the $e^+$ contamination. The muons continue undeflected through the
(uncharged) kicker and are focused on a 1.2-cm tall by 6.5-cm wide
aperture. The activated (charged) kicker induces a 36~mrad vertical
deflection, which causes the beam to be blocked at the aperture. In
2004, the average unkicked muon rate was limited to $2$~MHz;
approximately 10 muons were accumulated per cycle, of which 4
remained undecayed when $T_M$ began.

The kicker is described in detail in Ref.~\cite{kicker}.  Briefly,
it consists of two pairs of electrode plates biased to produce a
potential difference of up to $V_{K} = 25$~kV, with a virtual ground
at the midplane. Modulators, using series-connected MOSFETs
operating in push-pull mode, charge or discharge the plates. In
2004, a partial system achieved an average beam extinction of
$\varepsilon = 260$ with a 60~ns switching time~\cite{kickernow}.
During $T_M$, $V_K$ changed by less than 0.25~V.  A time dependence
of $V_K$ at this level, together with a voltage dependent
extinction, gives rise to a 3.5~ppm systematic error on the muon
lifetime.

The parity-violating correlation between the muon's spin orientation
and the emission direction of its decay positron can lead to a
systematic shift in the extracted lifetime, for the following
reasons: Suppose detector $A$ at position $(\theta,\phi)$ counts
positrons at the rate $N^{}_A(\vec{P}_{\mu})\exp(-t/\tau_{\mu})$,
where $\vec{P}_{\mu}$ is the polarization of the stopped muons. If
$\vec{P}_{\mu}$ varies with time because of relaxation or spin
rotation caused by
% external or internal ???
magnetic fields, so will $N^{}_A(\vec{P}_{\mu})$. A temporal
variation that is long compared to $\tau_{\mu}$ will manifest itself
as an unobserved distortion to the fitted lifetime of the detector.
The spin-related systematic uncertainty in $\tau_{\mu}$ is minimized
through both detector symmetry and target choice: the positron
detectors are arranged as a symmetric ball covering a large solid
angle, and every detector $A$ at position $(\theta,\phi)$ is
mirrored by another detector, $\bar{A}$ at
\mbox{$(\pi-\theta,\phi+\pi)$}. To the extent that the detector
pairs have the same geometrical acceptance and efficiency
%the sum of the coefficients,
$N_{A+\bar{A}} \equiv N^{}_A(\vec{P}_{\mu}) +
N_{\bar{A}}(\vec{P}_{\mu})$ is independent of the value of
$\vec{P}_{\mu}$, and therefore also its time variation. Finally, a
target possessing a high internal magnetic field is used so that the
muon spin precession period is $\ll \tau_{\mu}$.

\begin{figure}
\begin{center}
\includegraphics[width=\columnwidth]{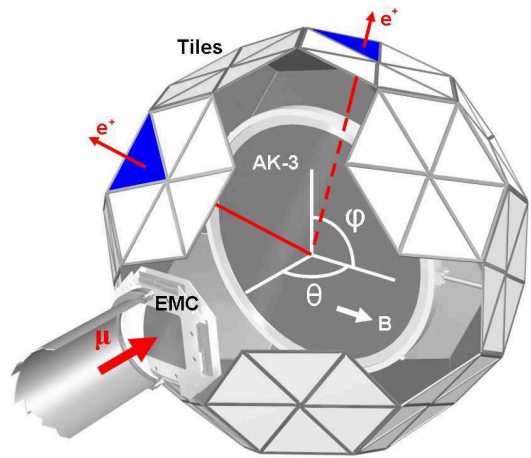}
\end{center}
\caption{Diagram of the experiment with several detector elements
removed. Muons enter through the beampipe vacuum window and traverse
the EMC and a helium bag (not shown) before stopping in the AK-3
target. Decay positrons are recorded by the coincidence of inner and
outer scintillators in one triangular segment; the outer
scintillators are visible. Two example decay trajectories are
shown.} \label{fig:detector}
\end{figure}

As depicted in Fig.~\ref{fig:detector}, the muon beam exits its
vacuum pipe through a 9.3-cm diameter, 76-$\mu$m-thick Mylar window,
then passes through a thin, high-rate multiwire entrance muon
chamber (EMC), which records the time and position of muons entering
the detector.  Roughly 1 in $10^4$ muons stop in the EMC. Their
spins precess in the field of a permanent magnet array, which has a
mean transverse field of 11~mT at the EMC center. The field
orientation was regularly reversed throughout data taking. The
region between the exit of the EMC and the target is spanned by a
helium-filled balloon (instead of air) to minimize muon stops and
scattering.

The stopping target is a 0.5-mm thick, 50-cm diameter disk of
Arnokrome$^{\rm TM}$~III (AK-3)~\cite{arnokrome}, having an internal
magnetic field of approximately 0.4~T, oriented transverse to the
muon spin axis. The field direction was reversed at regular
intervals. Dedicated $\mu$SR measurements~\cite{morenzoni} on an
AK-3 sample show an 18-ns oscillation period with a large initial
asymmetry that relaxes with a time constant of 14~ns.  These times
are considerably shorter than the accumulation period, $T_A$.  Using
the difference spectrum of counts from mirrored detectors versus
time; e.g., $N_{A - \bar{A}}$, the longer-term components are shown
to be negligible.

The decay positrons are recorded by 170 detector elements, each
consisting of an inner and outer layer of 3-mm-thick, BC-404 plastic
scintillator.  Each triangle-shaped scintillator is read edge-on
using a lightguide mounted at $90^{\circ}$, which is coupled to a
29-mm photomultiplier tube (PMT). The 170 elements are organized in
groups of six and five to form the 20 hexagon and 10 pentagon faces
of a truncated icosahedron (two pentagons are omitted for the beam
entry and symmetric exit). The distance from target center to an
inner scintillator is 40.5~cm. The total acceptance is $64\%$,
taking into account the reduction in the geometrical coverage of
$70\%$ from positron range and annihilation in the target and
detector materials.

A clip line reshapes the natural PMT pulse width to a full-width at
$20\%$ maximum of 9~ns. These signals are routed to
%Lecroy 3312 or 4413
leading-edge discriminators that have 10~ns output widths. On
average, a throughgoing positron gives a signal of
70~photoelectrons, producing a 400-mV pulse height. The data taking
was split almost evenly between periods of 80-mV and 200-mV
discriminator thresholds. The arrival time of a positron is measured
with respect to the kicker transition  by a CAEN V767 128-channel,
multihit TDC. An Agilent E4400B frequency synthesizer, operating at
approximately 190.2~MHz, serves as the master clock. Its absolute
frequency is accurate to $10^{-8}$ and its central value did not
change at this level over the course of the run. A clock step-down
and distribution system provides a 23.75~MHz square wave as the
input clock for each TDC.

The master clock frequency was given a concealed offset within
$250$~ppm from 190.2~MHz. The analyzers added a fitting offset to
$\tau_{\mu}$ when reporting intermediate results. Only after the
analysis was complete was the exact oscillator frequency revealed,
and the fitting offset removed, to obtain the lifetime.

The raw data consist of individual scintillator hit times for each
cycle. The kicker transition defines a common (global) $t=0$, using
the 1.32~ns resolution provided by the 32 subdivisions of the TDC
input oscillator period.  To avoid problems with differential
nonlinearities in the TDC clock period division circuit, coincidence
windows, artificial  deadtimes, and decay histogram bin widths were
always set at integer multiples of the undivided input clock period
(42~ns).

Events are missed if a positron passes through a detector during the
electronic or software-imposed deadtime following a recorded event
in the same detector. With peak rates in individual detectors of
7~kHz, the ``pileup'' probability for a 42-ns deadtime is $<
3\times10^{-4}$. If unaccounted for, this leads to a 67~ppm shift in
the fitted $\tau_{\mu}$. Pileup can be accommodated by including an
$\exp(-2t/\tau_{\mu})$ term in the fit function, but this doubles
the uncertainty on the fitted $\tau_{\mu}$.  Instead, an artificial
pileup spectrum, constructed from secondary hits occurring in a
fixed-width time window that is offset from a primary hit, is added
back to the raw spectrum, thus restoring, on average, the missing
hits. The procedure is repeated using a wide range of artificial
deadtime periods and offsets, and the corrected spectra all give
consistent lifetimes. The systematic uncertainty from this procedure
is 2~ppm.

Preliminary fits to the decay time spectra using the function $N(t)
= N_{0}\exp(-t/\tau_{\mu}) + B$ showed a common structure in the
residuals at early times, independent of experimental condition or
detector. The structure is caused by an intrinsic flaw in the TDC,
which does not lose events but can shift them in time by $\pm25$~ps.
This behavior was characterized in extensive laboratory tests using
white-noise and fixed-frequency sources, together with signals that
simulate the kicker transitions. For a fit start time of $t_{\rm
start}=1~\mu$s, the TDC response settles into a simple pattern that
can be described well by a modification of the decay time spectrum
by a factor
\begin{equation}
\Im(t') = 1+A\cos(2\pi t'/T + \delta) \exp(-t'/\tau^{}_{\rm TDC }),
\end{equation}
where $t'=t-t_{\rm start}$, and with typical values of $A =
5\times10^{-4}$, $\tau^{}_{\rm TDC }=600$~ns, and $T=370$~ns. A
spectrum of $10^{11}$ white-noise events was fit to a constant
function, modified by $\Im(t')$, achieving a good $\chi^2$ and
structureless residuals.

The function used to fit the decay time spectra is
\begin{equation}
N(t') = \Im(t')\cdot[ N_0 \exp(-t'/\tau_{\mu}) + B].
\label{eq:fitfunc}
\end{equation}
Because the clock frequency was blinded during the analysis, $T$ and
$\tau^{}_{TDC}$ in $\Im(t')$ could not be fixed in the fits. With
the clock frequency revealed, these parameters are found to be
consistent with the laboratory values.  In an important test of the
appropriateness of Eq.~\ref{eq:fitfunc}, the fitted $\tau_{\mu}$ is
found to be independent of the fit start time beyond the minimal
$t_{start} = 1.05~\mu$s. The systematic uncertainty for the TDC
response is 1~ppm.

Subsequent to the 2004 run, waveform digitizers (WFDs) replaced the
discriminator and TDC timing system. The WFDs establish the
stability of the PMT gain versus time during $T_M$. A gain change,
together with a fixed discriminator threshold (as in 2004), will
appear as a time-dependent efficiency. The analysis of the PMT gain
stability over a range of instantaneous rates indicates a systematic
effect of less than 1.8~ppm on $\tau_{\mu}$.

A powerful consistency test is performed by grouping data from
detectors having a common azimuthal angle $\phi$ (see
Fig.~\ref{fig:detector}), fitting each group independently, and
sorting the lifetime results by $\cos\theta$, where $\theta$ is the
polar angle. For the AK-3 data, the lifetime distribution is flat
over $\cos\theta$ ($\chi^{2}/\rm{dof}=17.9/19$). For data taken with
a 20-cm diameter sulfur target, surrounded by a permanent magnet
array, the same distribution is not flat ($\chi^{2}/\rm{dof} =
8.0$). The cause in the latter case is a higher fraction of muons
that miss the target and stop downstream along inner detector walls.
The SRIM program~\cite{sims} finds that $0.55\%$ of the muons miss
the smaller sulfur target, while only $0.07\%$ miss the AK-3 target.
In both cases, $\approx 0.11\%$ suffer large-angle scatters in the
EMC or backscatter from the target, stopping in upstream detector
walls. Simulated decay spectra for all ``errant'' muons---using the
muon stopping distribution predicted by SRIM and including detector
element acceptance, initial muon polarization, and relaxation---were
used to determine the expected distortion to the lifetime. The
procedure was tested against a special data set in which all
incoming muons were stopped in a plastic plate placed midway between
the EMC and the target. The distribution of lifetimes with
$\cos\theta$ was successfully reproduced. For the AK-3 target, a
distortion as large as 1~ppm could be expected. However, a negative
shift in the range of $4 - 12$~ppm is implied for sulfur,
principally owing to the $\sim8$ times larger fraction of downstream
muon stops. The uncertainty on a correction of this magnitude could
be as large as $100\%$, exceeding the 15~ppm statistical precision
of the sulfur data sample. Therefore, we chose not to use the sulfur
data in our reported results---a decision made prior to unblinding
the clock frequency---even though such a correction would bring the
fitted sulfur lifetime into excellent agreement with that found for
AK-3. We conservatively assign a 2~ppm systematic uncertainty to
this procedure for the AK-3 data set.

%\begin{figure}[t]
%\begin{center}
%\includegraphics[width=\columnwidth]{consistency.eps}
%\end{center}
%\caption{Lifetime deviations with respect to the central value
%versus experimental condition. Groups having the same symbols are
%independent subsets of the total data set. The shaded band
%represents the uncertainty for the full data set. Only statistical
%errors are given. \label{fig:consistency}}
%\end{figure}

Other small systematic uncertainties are listed in
Table~\ref{tab:systematics}. Data integrity checks indicate a small
fraction of hits ($6 \times 10^{-6}$) that may be duplicates of
earlier hits. When the duplicates are removed, $\tau_{\mu}$ shifts
upwards by 2~ppm. Since the status of these hits is uncertain, a
correction of +1~ppm is applied to $\tau_{\mu}$ with a systematic
uncertainty of 1~ppm.  A systematic error from queuing losses in the
TDC single-channel buffer is calculated to be less than 0.7~ppm. The
systematic error from timing shifts induced by previous hits in a
channel during the measurement period is less than 0.8~ppm.

The final result is based on a fit using Eq.~\ref{eq:fitfunc} to the
$1.8 \times 10^{10}$ events in the summed AK-3 spectra, giving
$$\tau_{\mu}({\rm MuLan}) = 2.197\,013(21)(11)~\mu{\rm
s~~~(11.0~ppm)}$$
with a $\chi^2/{\rm dof} = 452.5/484$.  The first error is
statistical and the second is the quadrature sum of the systematic
uncertainties in Table ~\ref{tab:systematics}.
Figure~\ref{fig:cycleplot} indicates the range of the fitted region
and the inset displays the residuals divided by their uncertainty in
units of standard deviations from the fit. The consistency of
$\tau_{\mu}$ was checked against  experimental conditions, including
detector, threshold, target and magnet orientation, extinction
factor, kicker voltage, and run number.
%Several checks
%are displayed in Fig.~\ref{fig:consistency}.  The low lifetime value
%corresponding to $V_k = 22$~kV includes
Only a sub-group of runs at the beginning of the production period
exhibited an anomalous lifetime compared to the sum. When all runs,
in groups of 10, are fit to a constant, a $\chi^{2}/\rm{dof} =
108/102$ is obtained, suggestive that the sub-group fluctuation was
statistical.

Our fitted lifetime is in excellent agreement with the world
average, $2.197\,03(4)~\mu$s, which is based on four previous
measurements~\cite{lifetimes}. The improved world average is
$$\tau_{\mu}({\rm W.A.}) = 2.197\,019(21)~\mu{\rm s~~~(9.6~ppm)}.$$
Assuming the standard model value of the Michel parameter $\eta =
0$, and light neutrinos, determines the Fermi constant
$$G_{F} = 1.166\,371(6) \times 10^{-5} ~{\rm GeV}^{-2} ~~~{\rm
(5~ppm)}.$$

In a companion Letter~\cite{MuCap}, a new determination of the
induced pseudoscalar coupling ${\textsl g}^{}_P$ is reported. It
depends mainly on a comparison of negative and positive muon
lifetimes, the latter quantity being reported here.

We acknowledge the generous support from PSI and the assistance of
its accelerator and detector groups. We thank W.\,Bertl,
J.\,Blackburn, K.\,Gabathuler, K.\,Dieters, J.\,Doornbos, J.\,Egger,
W.J.\,Marciano, D.\,Renker, U.\,Rohrer, R.\,Scheuermann,
R.G.\,Stuart and E.\,Thorsland for discussions and assistance. This
work was supported in part by the U.S. Department of Energy, the
U.S. National Science Foundation, and the John Simon Guggenheim
Foundation (DWH).

\begin{table}
\caption{Systematic uncertainties. \label{tab:systematics}}
\begin{tabular}{lc}
\hline\hline
Source    & \hspace{0.5cm} Size (ppm) \hspace{0.5cm}  \\
\hline
Extinction stability & 3.5     \\
Deadtime correction &  2.0    \\
TDC response  & 1.0     \\
Gain stability & 1.8    \\
Errant muon stops & 2.0     \\
Duplicate words (+1 ppm shift) & 1.0 \\
Queuing loss &  0.7    \\
Multiple hit timing shifts & 0.8     \\
\hline
Total  &  5.2    \\
\hline\hline
\end{tabular}
\end{table}


\begin{thebibliography}{99}
\bibitem{kickernow}Using the full system (four modulators instead of two),
$\varepsilon = 800$ is achieved with a 45~ns switching time.
\end{thebibliography}
\end{document}